\documentclass[aps,prd,twocolumn,superscriptaddress,preprintnumbers,floatfix,nofootinbib,balancelastpage]{revtex4-2}
\pdfoutput=1
\usepackage{slashed}
\usepackage[dvipsnames,table]{xcolor}
\usepackage{hyperref}
\usepackage{xspace}
\usepackage[tight]{subfigure}
\usepackage{amsmath}
\usepackage{amssymb}
\usepackage{amsfonts}
\usepackage{mathrsfs}
\usepackage{comment}
\usepackage{afterpage}
\usepackage{verbatim}
\usepackage{booktabs}
\usepackage{array}
\usepackage[title,titletoc]{appendix}
\usepackage{tabularx}
\usepackage{physics}
\usepackage{lipsum}
\usepackage{graphicx}
\usepackage{xcolor}
\usepackage{marginnote}

\newcommand{\as}{\alpha_s}
\newcommand{\sigbar}{\bar{\sigma}}

\newcommand{\dsigbar}{\mathrm{d}\sigbar}
\newcommand{\Nc}{N_c}
\newcommand{\nn}{\nonumber}
\newcommand{\TNP}{\mathrm{TNP}}

\begin{document}

\title{Robust estimates of theoretical uncertainties  at fixed-order in perturbation theory}
\author{Matthew A.~Lim}
\email{m.a.lim@sussex.ac.uk}
\affiliation{Department of Physics and Astronomy, University of Sussex, Sussex House, Brighton, BN1 9RH, UK}

\author{Rene Poncelet}
\email{rene.poncelet@ifj.edu.pl}
\affiliation{Institute of Nuclear Physics, ul. Radzikowskiego 152, 31--342 Krakow, Poland}

\begin{abstract}
Calculations truncated at a fixed order in perturbation theory are accompanied by an associated theoretical uncertainty, which encodes the missing higher orders (MHOU). This is typically estimated by a scale variation procedure, which has well-known shortcomings. In this work, we propose a simple prescription to directly encode the missing higher order terms using theory nuisance parameters (TNPs) and estimate the uncertainty by their variation. We study multiple processes relevant for Large Hadron Collider physics at next-to-leading and next-to-next-to-leading order in perturbation theory, obtaining MHOU estimates for differential observables in each case. In cases where scale variations are well-behaved we are able to replicate their effects using TNPs, while we find significant improvement in cases where scale variation typically underestimates the uncertainty.
\end{abstract}

\preprint{IFJPAN-IV-2024-15}

%%%%%%%%%%%%%%%%%%%%%%%%%%%%%%%%%%%%%%%%%%%%%%%%%%%%%%%%%%%%%%%%%%%%%%%%%%%%%%%%%%%%%%%%%%%%%%%%%%%%%%%%%%%%%%%%%%%%%%%%%%%%%%%%

\strut\hfill%\draftdate
\maketitle

%%%%%%%%%%%%%%%%%%%%%%%%%%%%%%%%
\section{Introduction}\label{sec:intro}
A long-standing problem in using renormalisable quantum field theories to make phenomeological predictions in perturbation theory is how one reliably estimates missing higher-order uncertainties (MHOU). The conventional method to gauge the size of these terms relies on the fact that, for a prediction truncated at a given order, the difference between predictions computed at different values of the unphysical renormalisation scale $\mu$ is of genuinely higher order. Typically, practitioners choose a `good' value of the central scale and take an envelope of scale variations as a measure of the uncertainty. One then expects the size of this uncertainty to reduce as one calculates successive orders. There remains, however, an arbitrariness in the choice of the central scale and in the size of the variations which one performs. In practice, one often finds that scale variations provide an underestimate of the theoretical uncertainty, due to e.g. the emergence of new partonic channels at higher orders, cancellations rendering corrections artificially small, etc. In addition, scale variations do not correlate uncertainties between bins of a differential distribution (or between multiple differential distributions) correctly; moreover, $\mu$ itself does not represent an actual parameter with a true value which the prediction depends upon, and thus cannot be relied upon to determine a true uncertainty. 

Various attempts have been made to place the MHOU estimation problem on a more stable footing. Ref.~\cite{Cacciari:2011ze} aimed to associate the MHOU with a Bayesian degree of confidence, based on the perturbative expansion of an observable. This was further refined in Ref.~\cite{Bonvini:2020xeo}, which aimed to better treat cases with large higher-order corrections. Both methods were scrutinised in Ref.~\cite{Duhr:2021mfd}, which also aimed to incorporate the dependence on the renormalisation and factorisation scales $\mu_R$, $\mu_F$ into Bayesian inference. Ref.~\cite{Ghosh:2022lrf} took a rather different approach, proposing an uncertainty estimate for an arbitrary process based on the scale variations of a set of QCD reference processes. 

Theory nuisance parameters (TNPs) were introduced in Ref.~\cite{Tackmann:2024kci} as a way to directly parameterise unknown higher-order terms. Recent applications have sought to use TNPs in conjunction with scale variations to estimate the uncertainty from unknown higher-loop ingredients at a given order in resummed perturbation theory~\cite{Dehnadi:2022prz,Ju:2023dfa,Cal:2024yjz}. This is highly convenient, since factorisation in Soft-Collinear Effective Theory dictates the structure of the resummed calculation to all orders. This information allows TNPs to be directly identified with unknown terms in the resummation, such as anomalous dimensions. In a similar vein, TNPs have been employed to obtain approximate N$^3$LO parton distribution functions by the MSHT collaboration~\cite{McGowan:2022nag}, where they have been used to parameterise both partially known PDF ingredients at N$^3$LO and completely unknown cross sections. TNPs thus serve as an appealing alternative to the scale variation procedure, even for calculations at fixed order~\cite{Tackmann:2024kci}. 

In this work, we apply TNPs to estimate missing higher-order uncertainties for fixed-order differential distributions. Unlike the resummed case, the all-order structure of a fixed-order calculation is \textit{a priori} unknown. This necessarily introduces a greater arbitrariness in exactly how the parameterisation is accomplished. We shall show, however, that using relatively simple ans\"atze we are able to produce uncertainty bands which show a good convergence between orders, replicate the effects of scale variation procedures in cases where these work well, and greatly improve on scale variation in cases where it fails. 

\begin{figure*}[t]
    \centering
    \includegraphics[width=1.0\linewidth]{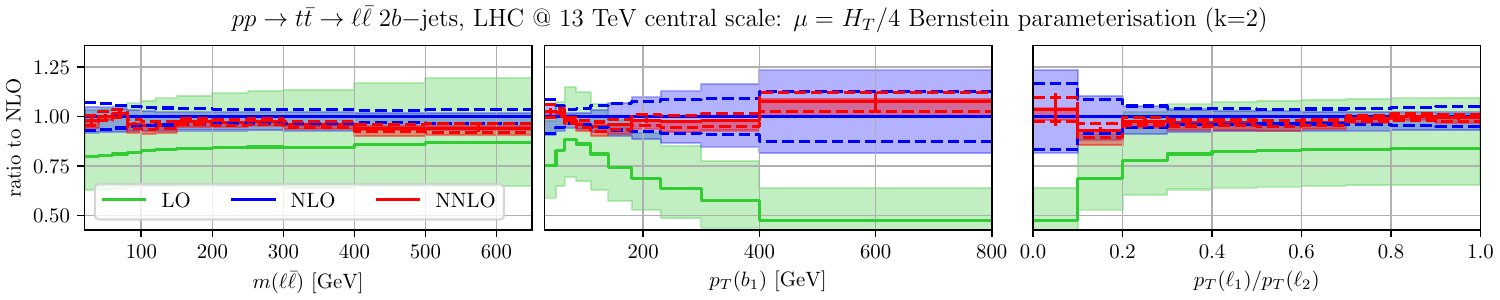}
    \caption{Comparison between MHOU estimates from TNP and scale variation for differential distributions in  $t\bar{t}$ production, with QCD corrections to the production and decay processes treated separately and combined in the narrow width approximation. From left to right, we show the lepton pair invariant mass $m(\ell\bar{\ell})$, the transverse momentum of the hardest $b$-jet $p_T(b_1)$ and the ratio of the lepton transverse momenta $p_T(\ell_1)/p_T(\ell_2)$. We take $\mu=H_T/4$ as our central scale choice and use a Bernstein parameterisation. The solid lines represent the LO (green), NLO (blue) and NNLO (red) central predictions and the corresponding bands of the estimated MHOU from scale variations, the dashed bands those from TNPs. The vertical lines indicate statistical uncertainties. We do not associate a TNP uncertainty with the LO prediction -- see text for details.}
    \label{fig:ttbar-HT4}
\end{figure*}

\begin{figure*}[t]
    \centering
    \includegraphics[width=1.0\linewidth]{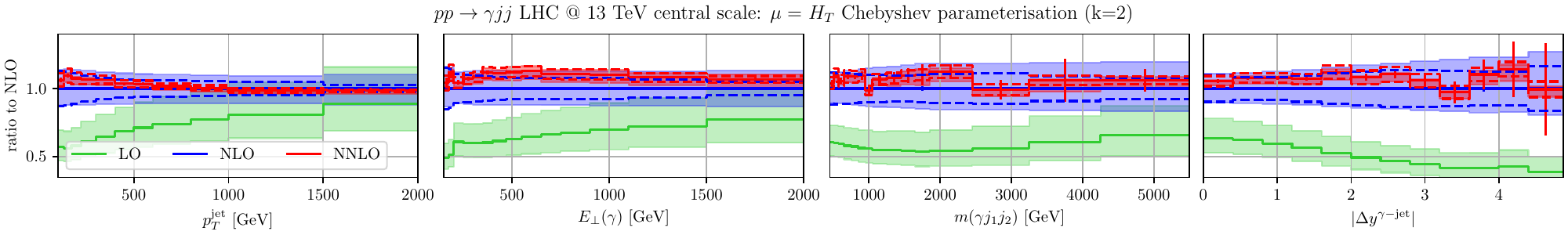}
    \caption{As in fig.~\ref{fig:ttbar-HT4}, but for $\gamma jj$ production. We take $\mu = H_T$ as our central scale choice and use a Chebyshev parameterisation. From left to right we show the hardest jet transverse momentum $p_T^{\rm jet}$, the photon transverse energy $E_T(\gamma)$, the invariant mass of the system $m(\gamma j_1 j_2)$ and the absolute rapidity difference between photon and hardest jet $|\Delta y^{\gamma- \rm{jet}}|$.}
    \label{fig:ajj-HT}
\end{figure*}

\section{Implementation}\label{sec:impl}
In contrast to resummed perturbation theory, given a computation at a certain fixed order we can infer very little about the structure of higher-order corrections \textit{a priori}. We must therefore introduce some parameterisation of the missing higher orders. We begin by writing the expansion of a differential cross section in the strong coupling $\as$ as
\begin{align}
    \mathrm{d}\sigma 
    %&= \overbrace{\overbrace{\dsig^{(0)}}^{\dsig_{\mathrm{LO}}} + \dsig^{(1)}}^{\dsig_{\mathrm{NLO}}}+ \dsig^{(2)}+\dots \nn \\
    &= \as^n \Nc^m \,\dsigbar^{(0)} + \as^{n+1} \Nc^{m+1}\,\dsigbar^{(1)}\nn \\
    & \qquad \qquad \qquad \qquad +\as^{n+2} \Nc^{m+2} \, \dsigbar^{(2)}+\dots \nn \\
    &= \as^n\Nc^m \,\dsigbar^{(0)} \bigg[1+\as\Nc\left(\frac{\dsigbar^{(1)}}{\dsigbar^{(0)}}\right) \nn \\
    & \qquad \qquad \qquad \qquad + \as^2\Nc^2\left(\frac{\dsigbar^{(2)}}{\dsigbar^{(0)}}\right)+\dots\bigg]
    \label{eq:expansion}
\end{align}
where we have abbreviated the differential cross section as $\mathrm{d}\sigma$ and factored out the leading power of the number of colours $N_c$ in each term.

At a given order $N$ in fixed-order perturbation theory, the term of order $N+1$ in eq.~\ref{eq:expansion} provides an estimate of the theoretical uncertainty associated with truncating the series. It is this term which we wish to model using nuisance parameters. To do this, we use information from available lower orders of the calculation. Specifically, we assume that the coefficients of $\alpha_s N_c$ in the square brackets of equation~\ref{eq:expansion} are simple functions of the kinematics such that they can be modelled by a polynomial with coefficients of order 1 multiplying a lower-order term. We find this empirically to be true for a range of processes and distributions. This motivates us to write the order $N+1$ estimate as
\begin{align}
    \frac{\dsigbar^{(N+1)}_{\TNP}}{\dsigbar^{(0)}} = \sum _{j=1}^N f^{(j)}_k\Big(\vec{\theta},x\Big) \left(\frac{\dsigbar^{(j)}}{\dsigbar^{(0)}}\right)
    \label{eq:general}
\end{align}
where $f_k$ is a polynomial of degree $k$ in a (suitably rescaled) kinematic variable $x$ and the $\theta_i$ are the TNPs.\footnote{In the notation of Ref.~\cite{Tackmann:2024kci}, this is then an N$^{m+1}$LO model for a fixed value of the scale $\mu$.} We have experimented with 
different choices of polynomial basis, and present results using Bernstein polynomials:
\begin{align}f_k^B\Big(\vec{\theta},x\Big) &=
\sum^k_{i=0} \theta_i  \bigg(\begin{array}{c} k \\ i \end{array}\bigg) x^{k-i} (1-x)^i \,
    \label{eq:poly}
\end{align}
defined in the range $x\in[0,1]$ as well as for Chebyshev polynomials of the first kind,
\begin{align}
f_k^C\Big(\vec{\theta},x\Big) &=\frac{1}{2}\sum^k_{i=0} \theta_i  T_i(x) \, ,
\end{align}
defined in the range $x\in[-1,1]$. The choice of the integer $k$ modulates the amount of shape variation which is permitted. In this work we always use $k=2$. \footnote{Choosing $k=0$ would mirror closely the procedure followed in Ref.~\cite{McGowan:2022nag} at N$^3$LO. In principle, in cases where one has additional information about the process it would be possible to refine eq.~\ref{eq:general} further. For example, for processes where new channels appear at higher orders, one could decompose each $\dsigbar^{(i)}$ according to the partonic initial states $\dsigbar^{(i)}_{gg}$, $\dsigbar^{(i)}_{q\bar{q}}$, etc. and treat each separately. Since we find our results are already consistent over a range of processes and distributions, we do not attempt this here.}

 We remark that while the parameterisation in eq.~\ref{eq:general} is of course arbitrary, all choices should be equivalent in the limit $k\to \infty$. The arbitrariness actually enters as soon as we choose to truncate the polynomial expansion, inducing dependence both on the exact parameterisation and the truncation order. A discussion of parameterisation sensitivity is given in Ref.~\cite{Tackmann:2024kci}. Empirically, we find that employing our procedure at NLO gives a good approximation to the known NNLO term for $\theta_i\sim \mathcal{O}(1)$, which we use as justification to extend this one order higher to gauge the size of missing N$^3$LO terms. The point we wish to stress, however, is that despite any residual ambiguities, we believe this still represents an improvement on the scale variation procedure. For some choice of the $\theta_i$ in eq.~\ref{eq:general} and $k=2$, we find that it is always possible to construct at least a fair approximation to the next unknown order. In contrast it is frequently the case that, at a given order, no choice of $\mu$ will reproduce the next order. We will demonstrate this explicitly in sec.~\ref{sec:results}.

\begin{figure*}[t]
    \centering
    \includegraphics[width=1.0\linewidth]{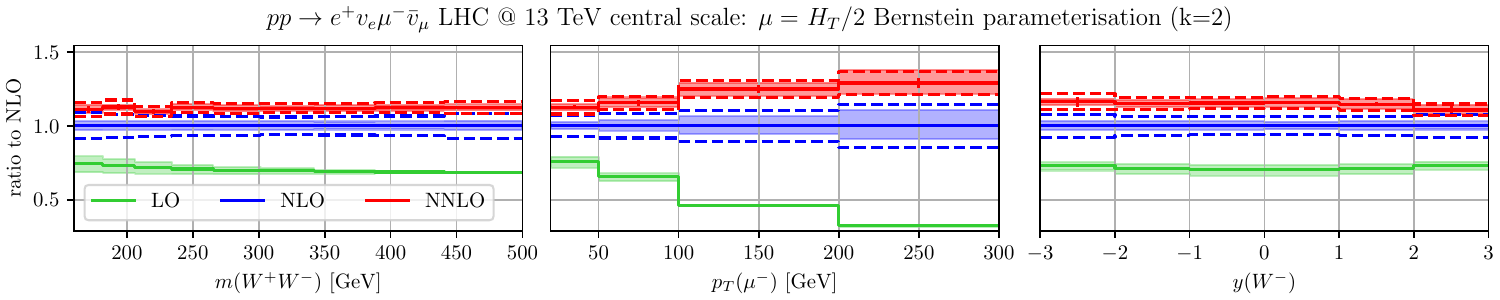}
    \caption{As in fig.~\ref{fig:ttbar-HT4}, but for off-shell, fully decayed $WW$ production. We take $\mu = H_T/2$ as our central scale choice and use a Bernstein parameterisation. From left to right we show the pair invariant mass $m(W^+W^-)$, the transverse momentum of the muon $p_T(\mu^-)$ and the rapidity of one boson $y(W^-)$.}
    \label{fig:ww-HT2}
\end{figure*}

\begin{figure*}[t]
    \centering
    \includegraphics[width=0.5\linewidth]{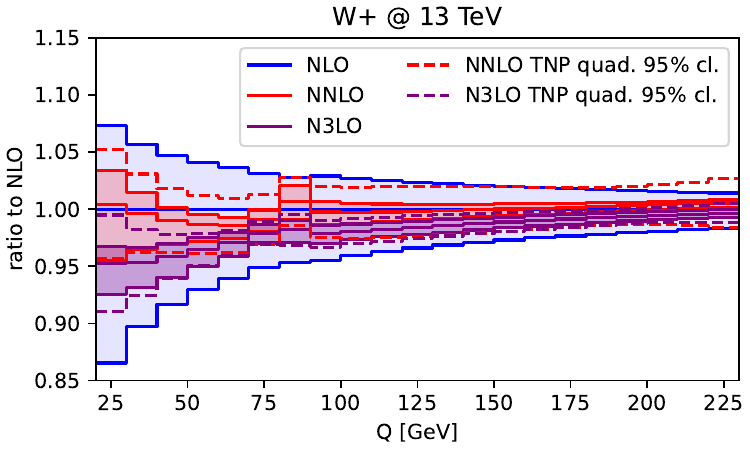}%
    \includegraphics[width=0.5\linewidth]{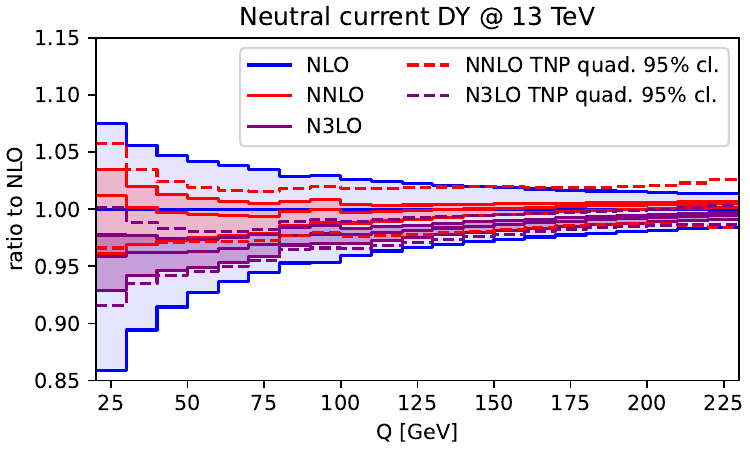}%
    \caption{Comparison between scale and TNP uncertainties for charged-current ($W^+$) and neutral-current Drell-Yan ($Z/\gamma$) through next-to-next-to-next-to-leading order. The coloured bands indicate the uncertainties from 7-point scale variations, and dotted lines the quadrature combination of the TNP uncertainties at a 95\% CL.}
    \label{fig:DYN3LO}
\end{figure*}

We do not attempt to construct a TNP model for the LO uncertainty. In this case, there is no lower-order information upon which we could reasonably base our ansatz. Given that, thanks to automation, NLO has become the de facto minimum standard for the majority of processes, we do not consider this to be a major shortcoming. 

We note that strict adherence to the philosophy espoused in Ref.~\cite{Tackmann:2024kci} would dictate that our model be independent of the size (if not the shape) of lower-order corrections. The motivation for this is to avoid significant underestimation of the MHOU when the lower-order terms happen by chance to be smaller than might be na\"ively expected. While this is certainly a desirable property in principle, in practice we have found it difficult to achieve, even after factoring in known structures (flavour decomposition of partonic channels, etc.). Since our aim here is to provide a prescription which can be applied generally and to a range of processes, we have conceded on this particular tenet\footnote{The closest analogue in Ref.~\cite{Tackmann:2024kci} would be strategy 3 of section 4.3. In that case, however, the author is explicit about the normalisation required to make the estimate independent of the size of lower order terms. In using this information, our approach is instead closer to that adopted in Ref.~\cite{McGowan:2022nag}.}. We note that in so doing, we do not observe any instances where we believe the MHOU has been significantly underestimated (in contrast to results obtained using normal scale variations). While we do not exclude the possibility of achieving a parameterisation which does not utilise this size information, its form is likely to be specific to a given interest case, given that it depends on a complex interplay of factors (e.g. partonic channels, the definition of fiducial regions, isolation criteria).

Another criticism that one might reasonably level is that we are still required to make a choice for the central value of our scales, which remains arbitrary. By examining different central choices, we have found that our uncertainty estimates remain consistent, regardless of what choice one makes. The underlying reason for this is that our construction naturally overcomes `poor' scale choices, in the sense that these result in large $K$-factors which then inflate the TNP uncertainty. This issue is also discussed in detail in Ref.~\cite{Tackmann:2024kci}.

\begin{figure*}[t]
    \centering
    \includegraphics[width=1.0\linewidth]{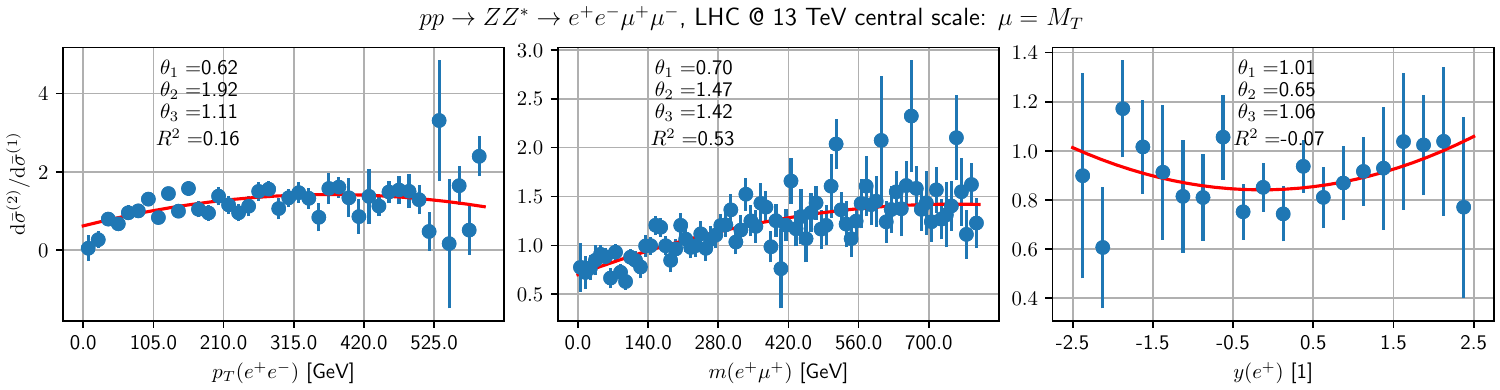}
    \caption{Fit of the NLO uncertainty estimate using Bernstein polynomials (red) to the true NNLO result (blue) for the process $pp\to ZZ^*\to e^+e^-\mu^+\mu^-$. For each kinematic distribution the corresponding fit values of the nuisance parameters are shown.}
    \label{fig:ZZfit}
\end{figure*}

\section{Results}
\label{sec:results}
We obtain our (N)NLO cross sections from published results obtained within the STRIPPER framework~\cite{Czakon:2010td, Czakon:2011ve,Czakon:2014oma,Czakon:2019tmo}: a full list of included processes can be found in Tab.~\ref{tab:processes}. We study both NLO uncertainties (where NNLO corrections are known) and NNLO uncertainties (where N$^3$LO corrections are at present unknown). We explicitly present results for three processes, namely $pp \to t\bar{t} \to b\bar{b}\ell\bar{\ell}$,  $pp\to \gamma jj$  and $pp \to W^+W^-\to \ell\bar{\ell}\nu\bar{\nu}$, all defined in appropriate fiducial regions. More processes are available using a Jupyter notebook interface to HighTEA at Ref.~\cite{highteacolab}. Each case has a distinct behaviour with respect to scale variations. The first and second processes feature coloured final states and large higher-order corrections, where typically the leading-order scale band does not contain the NLO result. The third example is pathological with respect to scale variations, and the NNLO uncertainty band lies completely outside that of the NLO. This allows us to test our procedure in different scenarios. For the original NNLO computations of each process, see Refs.~\cite{Behring:2019iiv,Gehrmann:2014fva, Grazzini:2016ctr, Badger:2023mgf}.

For each process and distribution we consider, we compute the central curve using a fixed scale choice. We compute uncertainty estimates by summing in quadrature variations by $\pm 1$ of each nuisance parameter separately. This is appropriate if the parameters are independently and normally distributed, a matter which we shall address in the following. We show results using both Bernstein and Chebyshev parameterisations depending on the process -- we have verified that the results are very similar, no matter which is used.

We begin by discussing results for the $t\bar{t}$ process with a central scale choice of $H_T/4$. In Fig.~\ref{fig:ttbar-HT4} we show the invariant mass of the lepton pair $m(\ell\bar{\ell})$, the transverse momentum of the hardest $b$-jet $p_T(b_1)$ and the ratio of the lepton transverse momenta $p_T(\ell_1)/p_T(\ell_2)$. From the scale variation plots we see that the LO uncertainty is generally underestimated but that beyond this order, scale variations appear to estimate the MHOU well. The TNP approach reproduce this behaviour, with marginally smaller estimates of the NLO uncertainty. 

\begin{figure*}[t]
    \centering
    \includegraphics[width=1.0\linewidth]{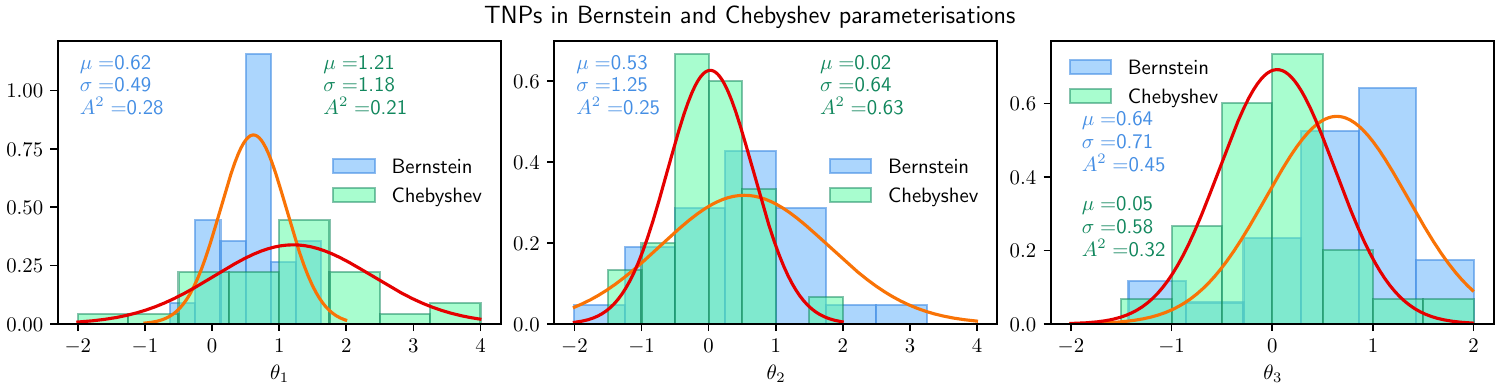}
    \caption{Distribution of the values of the nuisance parameters $\theta_i$ in Bernstein and Chebyshev parameterisations at NNLO. For details of the data points included, see text. The mean and standard deviation of a fit Gaussian distribution are indicated on the plot, as is the Anderson-Darling test statistic (critical value 0.712 for $p=0.05$). Binning is purely illustrative.}
    \label{fig:TNPdistributions}
\end{figure*}

Next, we examine in Fig.~\ref{fig:ajj-HT} the case of $\gamma jj$ production at a central scale of $H_T$. Examining first the lower plots, we note that, as in the $t\bar{t}$ case, the LO scale band does not contain the nominal prediction at NLO for any of the four distributions presented. Thereafter, scale variations again seem to estimate the MHOU well, as does the TNP approach. We stress here that the effect of the TNP variations is not only an increase in the size of the uncertainties, but also a modification of their shape. This is most clearly visible at NNLO in the $p_T^j$ and $|\Delta y^{\gamma-j}|$ distributions, where the scale variations give an asymmetric uncertainty while the TNP variation is more even.

In Fig.~\ref{fig:ww-HT2} we turn to off-shell $WW$ production, a process notorious for the poor behaviour of traditional methods in estimating MHOU terms. In this preliminary study, we do not include the effect of the loop-induced $gg$ channel at NNLO. From the lower scale variation plots we see that the uncertainty at each order is completely underestimated. This is a result of so-called `giant' $K$-factors, which arise from NLO corrections in which a vector boson is emitted off a final-state parton~\cite{Bauer:2009km,Denner:2009gj,Rubin:2010xp}. The size of these corrections is exacerbated by fiducial cuts. In the case of the invariant mass distribution, the LO band is in fact smaller that the NLO which is of a similar size to the NNLO scale variation. In none of the distributions do the scale bands even overlap. In contrast, while the process is clearly challenging even for the TNP approach, a much more sensible behaviour is observed. The TNP estimates decrease in size as one moves from NLO to NNLO, and the NLO error in particular seems to be much better estimated. 

Finally, we discuss the case of neutral and charged current Drell-Yan production. This is a rare case where fully-differential corrections in QCD are currently known through N$^3$LO~\cite{Chen:2021vtu,Chen:2022cgv} -- indeed, in Refs.~\cite{Duhr:2020sdp, Duhr:2020seh} the first predictions at this order for the invariant mass distribution of the dilepton pair were presented. It was observed that, beginning at N$^3$LO, the convergence of the perturbative series as measured by scale variation begins to deteriorate. While this is partly due to the effect of using NNLO PDFs (as opposed to aN$^3$LO~\cite{PDFforumtalk}), it is nevertheless interesting to examine whether TNPs provide an uncertainty estimate which suggests greater consistency between perturbative orders. In Fig.~\ref{fig:DYN3LO} we show predictions for the invariant mass distribution at various orders in perturbation theory, for both charged and neutral current Drell-Yan production, and compare scale variations with (Chebyshev) TNP uncertainty estimates at the 95\% confidence level.~\footnote{This corresponds to varying the TNPs by 2, rather than 1 as done elsewhere. We wish to stress that, given the statistical interpretation of the TNPs which we will demonstrate in the next section, this genuinely corresponds to a $2\sigma$ uncertainty -- it is thus \textit{not} analogous to a variation of $\mu$ by a factor of 4.} Results have been obtained using the public code \texttt{n3loxs}~\cite{Baglio:2022wzu}. A clear improvement of the description of the NNLO uncertainty in the TNP case with respect to the scale variation case is visible, particularly for $Q>100~\mathrm{GeV}$ where the uncertainties now enclose the N$^3$LO result. We are thus able to employ our method, without modification and in a simple and straightforward manner, to resolve an important discrepancy at the highest available order in perturbation theory. 

As an \textit{a posteriori} check on our procedure, we have performed fits of our NLO uncertainty estimate in cases where the NNLO distribution is known. An example of such a fit for the process $pp\to ZZ^*\to e^+e^-\mu^+\mu^-$ is shown in Fig.~\ref{fig:ZZfit}. This confirms that our parameterisation is able to fit the NNLO data with values of $\theta_i\sim \mathcal{O}(1)$. 

At this point, a comment is in order. It is true that, without knowledge of the functional form of the higher-order corrections (which one possesses, for example, in a resummed calculation), it is not possible to capture the correct bin-to-bin correlations of the uncertainties in a given differential distribution. This is only possible to the extent that the parameterisation in eq.~\ref{eq:general} provides a good approximation of the true higher-order behaviour. Given that this `true' behaviour is in practice always accompanied by a finite Monte Carlo error, the quality of the fits in Fig.~\ref{fig:ZZfit} demonstrate that this is indeed the case: that is, the parameterisation describes correlations as well as the integration error allows. It is of course possible that by increasing statistics, one might find that a polynomial with $k=2$ no longer describes the functional form well, and in fact underfits. In this case, one could always extend the parameterisation by choosing $k>2$.~\footnote{We have studied the effect of truncation uncertainties using a toy model and simulated noise. Our expectation is that the induced truncation error is currently small compared to the nominal uncertainty estimate.} Alternatively, it may be the case that by including too many TNPs one overfits. The treatment here depends on the exact parameterisation chosen. In the Chebyshev case, the orthogonality of the polynomial basis results in higher order coefficients naturally becoming smaller. In the Bernstein case, correlations between the TNPs are introduced which effectively set the coefficients of higher order monomials to zero. Care must therefore be taken when combining uncertainties or using them in a fit.

We have repeated the fit for a range of processes and distributions, which are listed in Tab.~\ref{tab:processes}. In selecting the kinematic distributions to be included, we have endeavoured to make choices which are quasi-independent -- for example, choosing a rapidity distribution, an invariant mass and a transverse momentum rather than two invariant masses. In this way, we have tried to avoid double-counting information arising from strongly correlated distributions. This also restricts the amount of information we are able to include from each process -- for $2\to 2$ processes we limit ourselves to three distributions, while more can be included in the $2\to3$ cases due to the greater kinematic freedom they afford. In total, we are able to include 30 data points. 

Results for the fit values of the nuisance parameters are shown in figs.~\ref{fig:TNPdistributions}, for both Bernstein and Chebyshev parameterisations. We observe that the resulting distributions appear to be Gaussian, with values of the mean and standard deviation of order 1. Test statistics obtained using the Shapiro-Wilk and Anderson-Darling methods show no significant deviation from normality at the $p=0.05$ level. This is a strong justification for our procedure, at least at NLO -- a similar validation at NNLO is unfortunately dependent on a greater availability of fully differential calculations at N$^3$LO.

\begin{table*}[]
\setlength{\tabcolsep}{4.pt}
 \renewcommand*{\arraystretch}{1.4}
\centering
\begin{tabular}{c c c c c c}
Process\,  & $\sqrt{s}/\mathrm{TeV}$              & Scale             & PDF             & Distributions & Reference \\
\hline \hline
$pp\to H$ (full theory)                      & 13 & $m_H/2$           &\texttt{NNPDF3.1}& $y_H$                                                                   & \cite{Czakon:2023kqm, Czakon:2024ywb}\\

$pp\to ZZ^* \to e^+ e^- \mu^+ \mu^-$         & 13 & $M_T$             &\texttt{NNPDF3.1}& $M_{e^+\mu^+}$, $p_{T}^{e^+e^-}$, $y_{e^+}$                               & \cite{Poncelet:2025xxx}\\

$pp\to WW^*\to e \nu_e \mu \nu_\mu$           & 13 & $m_W$             &\texttt{NNPDF3.1}& $M_{WW}$, $p_T^{\mu^-}$, $y_{W^-}$                                      & \cite{Poncelet:2021jmj} \\
$pp\to (W\to \ell \nu)+c$                    & 13 & $E_T+p_T^c$       &\texttt{NNPDF3.1}& $p_T^\ell$ ,$|y_\ell|$                                                 & \cite{Czakon:2022khx,CMS:2023aim}\\
$pp\to t\bar{t}$                             & 13 & $H_T/4$           &\texttt{NNPDF3.1}& $M_{t\bar{t}}$, $p_T^t$, $y_t$                                          & \cite{Czakon:2020qbd,CMS:2024ybg} \\
$pp\to t\bar{t} \to b\bar{b} \ell \bar{\ell}$& 13 &$H_T/4$            &\texttt{NNPDF3.1}& $M_{\ell\bar{\ell}}$, $p_T^{b_1}$, $p_{T,\ell_1}/p_{T,\ell_2}$          & \cite{Czakon:2020qbd,CMS:2024ybg} \\
$pp\to \gamma\gamma$                         & 8  & $M_{\gamma\gamma}$&\texttt{NNPDF3.1}& $M_{\gamma\gamma}$, $p_T^{\gamma_1}$, $y_{\gamma\gamma}$                & \cite{Czakon:2023hls}\\
$pp\to \gamma\gamma j$                       & 13 & $H_T/2$           &\texttt{NNPDF3.1}& $M_{\gamma\gamma}$, $p_T^{\gamma\gamma}$, $\cos\phi_{\mathrm{CS}}$,
                                                                                          $|y_{\gamma_1}|$, $\Delta \phi_{\gamma\gamma}$                            & \cite{Chawdhry:2021hkp} \\
$pp\to jjj$                                  & 13 & $\hat{H}_T$       &\texttt{MMHT2014}& {\scriptsize $\mathrm{TEEC}$ with ${H_{T,2} \in [1000,1500), [1500,2000), [3500,\infty) ~\mathrm{GeV}}$} & \cite{Czakon:2021mjy, Alvarez:2023fhi} \\
$pp\to \gamma jj$                            & 13 & $H_T$             &\texttt{NNPDF3.1}& $M_{\gamma jj}$, $p_T^j$, $|y_{\gamma-\text{jet}}|$, $E_{T,\gamma}$     & \cite{Badger:2023mgf} \\
\end{tabular}
\caption{Processes and NNLO distributions included in meta-analysis.}
\label{tab:processes}
\end{table*}

\section{Conclusions}
In this work we have proposed a simple prescription to estimate missing higher-order uncertainties for fixed-order differential distributions using theory nuisance parameters, which provides an appealing alternative to the scale variation paradigm. Our prescription is easily implemented at any order in perturbation theory and can be applied to any process and any inclusive observable.

We have made a Juypter notebook which implements the procedure discussed here publicly available~\cite{highteacolab}, making use of the HighTEA interface to NNLO calculations~\cite{Czakon:2023hls}. HighTEA acts as a user-friendly repository for NNLO events, allowing the user to request desired kinematic distributions, PDF sets and binnings. At present, a number of processes have been implemented, including cases where the TNP approach provides a dramatic improvement on scale variations (e.g. $\gamma\gamma$ production).

Our method could also be used in the context of electroweak corrections. These present a problem to scale variation procedures due to the weak dependence on $\mu$ of the electroweak coupling. A nuisance parameter approach, on the other hand, would not suffer from this problem and could thus improve uncertainty estimates in a consistent fashion. 

In practical applications, one is often interested in capturing not only the correlations in uncertainties between bins of a given kinematic distribution as is done here, but also between different distributions for a given process. Combining information in this way can be used to extract Standard Model parameters precisely, see e.g. ref.~\cite{Cridge:2023ztj, Cridge:2024exf}. Such a combination would require a multi-dimensional TNP treatment, which parameterises the joint dependence in multiple variables $x_i$. A simple attempt at this could for example use products of single-differential parameterisations. We leave a more detailed investigation of this topic to future work. Achieving the correct correlation model for differential distributions between processes would instead involve a significant departure from the approach we have adopted here and will require substantially new ideas.

We would encourage practitioners of higher-order calculations to investigate the method we have presented in this work, alongside normal scale variations. This would test how generalisable our method is, and potentially result in better convergence. Crucially, the simplicity of our proposal means that performing these estimates comes at no additional cost -- the uncertainty can be calculated using results one would already have obtained.

\section*{Acknowledgements}

We are grateful to Frank Tackmann and Thomas Cridge for discussions on this topic and for comments on the manuscript. We thank Tom Schellenberger for preparing the numerical results for Higgs production. This work was initiated at the Aspen Center for Physics, which is supported by National Science Foundation grant PHY-2210452 and by a grant from the Simons Foundation (1161654, Troyer). MAL is supported by the UKRI guarantee scheme for the Marie Sk\l{}odowska-Curie postdoctoral fellowship, grant ref. EP/X021416/1. RP thanks the CERN Theoretical Physics Department for hospitality while part of this work was carried out.

%%%%%%%%%%%%%%%%%%%%%%%%%%%%%%%%
\bibliographystyle{apsrev4-1}
\bibliography{lit}

\end{document}